\documentclass[twocolumn,reprint,aps,superscriptaddress,pra]{revtex4-1}
\usepackage[utf8]{inputenc}
\usepackage{color}
\usepackage{amsmath}
\usepackage{amssymb}
\usepackage{amsbsy}
\usepackage{bbold}
\usepackage{graphicx}
\usepackage{xfrac}
\usepackage{physics}
\usepackage{textcomp}
\usepackage[]{todonotes}
\usepackage[unicode=true,
 bookmarks=false,
 breaklinks=false,pdfborder={0 0 1},backref=false,colorlinks=true]
 {hyperref}
\hypersetup{
 linkcolor=blue,citecolor=blue,urlcolor=blue,linkcolor=blue,citecolor=blue,urlcolor=blue}

\makeatletter

\usepackage{braket}

\definecolor{mygreen}{rgb}{0,0.5,0}
\definecolor{myblue}{rgb}{0,0,0.75}
\definecolor{mymagenta}{cmyk}{0,1,0,0.12}

\makeatother

\begin{document}

\title{Large-area quantum-spin-Hall waveguide states in a three-layer topological photonic crystal heterostructure}

\author{Zhihao Lan}
\affiliation{Department of Electronic and Electrical Engineering, University College London, Torrington Place, London, WC1E 7JE, United Kingdom}
\author{Menglin L. N. Chen}
\affiliation{Department of Electronic and Information Engineering, The Hong Kong Polytechnic University, Kowloon, Hong Kong, People's Republic of China}

\author{Jian Wei You}
\affiliation{State Key Laboratory of Milimeter Waves, Institute of Electromagnetic Space, Southeast University, Nanjing, China}

\author{Wei E. I. Sha}
\affiliation{Key Laboratory of Micro-nano Electronic Devices and Smart Systems of Zhejiang Province, College of Information Science and Electronic Engineering, Zhejiang University, Hangzhou 310027, China}

\begin{abstract}
Topological photonic edge states are conventionally formed at the interface between two domains of topologically trivial and nontrivial photonic crystals. Recent works exploiting photonic quantum Hall and quantum valley Hall effects have shown that large-area topological waveguide states could be created in a three-layer topological heterostructure that consists of a finite-width domain featuring Dirac cone sandwiched between two domains of photonic crystals with opposite topological properties. In this work, we show that  a new kind of large-area topological waveguide states could be created employing the photonic analogs of quantum spin Hall effect. Taking the well-used Wu-Hu model in topological photonics as an example, we show that sandwiching a finite-width domain of photonic crystals featuring double Dirac cone between two domains of expanded and shrunken unit cells could lead to the emergence of large-area topological helical waveguide states distributed uniformly in the middle domain. Importantly, we unveil a power-law scaling regarding to the size of the bandgap within which the large-area helical states reside as a function of the width of the middle domain, which implies that these large-area modes in principle could exist in the middle domain with arbitrary width. Moreover, pseudospin-momentum locking unidirectional propagations and robustness of these  large-area waveguide modes against sharp bends are explicitly demonstrated. Our work enlarges the photonic systems and platforms that could be utilized for large-area-mode enabled topologically waveguiding.
\end{abstract}

\maketitle

\textit{Introduction.---} While topological states of matter were originally discovered in solid state electronic materials \cite{TI_review1,TI_review2}, the single-particle topological band theory does not rely on the fermionic nature of electrons and their Fermi-Dirac statistics. Haldane and Raghu \cite{QHE_PRL08_Haldane,QHE_PRA08_Haldane} made the first effort to extend the quantum Hall physics of two-dimensional electron gases under strong magnetic fields to photons in magneto-optic photonic crystals, where they predicted the existence of  chiral electromagnetic states whose energy can only propagate in a single direction. Such unidirectional backscattering-immune topological electromagnetic states were confirmed in experiments shortly thereafter \cite{QHE_Nature09_Wang}. As magneto-optical effects in general are weak in near-infrared and visible light regions, time-reversal symmetry preserving topological photonic systems without magneto-optical materials have later been proposed based on other members of the quantum Hall related states, such as quantum spin Hall \cite{QSH_NM13_Khanikaev,QSH_NC14_Chen,QSH_PRL15_Ma,QSH_15PRL_Wu,QSH_PNAS16_He} and quantum valley Hall \cite{QVH_NJP16_Ma,valley_review1,valley_review2} as well as higher-order topological phenomena \cite{Xie18PRB_corner,Kim20light_review,Xie21NRP_review}. The research field of topological photonics \cite{review_NP14,review_NP17,review_RMP19,review_RevPhys22} has not only deepened our understanding about topological physics in bosonic systems but also found many interesting applications.

The conventional way to create topological edge states in photonic systems is based on an interface between two photonic crystals of different topological properties. The edge states constructed in this way typically are tightly confined around the interface and decay exponentially away from it. Moreover, due to the wave nature of light, tunneling and coupling of topological edge states at different interfaces can also give rise to interesting physics \cite{Chen19PRB_magneticLayer,Jiang19AO_layered,Chen20IEEE_layered,Yang22OE_layered}. Recently, interesting large-area topological waveguide states were demonstrated in three-layer heterostructures in both photonic quantum Hall \cite{Wang21PRL_large,Qu22PRB_large} and quantum valley Hall \cite{Chen21ACSpho_largevalley} systems. In such topological heterostructures, a domain of photonic crystal featuring gapless Dirac cone dispersion is sandwiched between two domains of photonic crystals with opposite topological properties and the resulting topological waveguide states have field amplitudes distributed almost uniformly in the middle domain. In \cite{Wang21PRL_large}, one-way large-area topological waveguide states were created in magnetic photonic crystals with opposite gap Chern numbers and the optical forces of such waveguide states could be used for  particle sorting and manipulation \cite{Qu22PRB_large}. In \cite{Chen21ACSpho_largevalley}, a photonic crystal with Dirac points was sandwiched between two valley photonic crystals with opposite valley-Chern numbers and large-area valley-locked waveguide states were observed in experiments. Such large-area topological waveguide states with a finite width have a high capacity for energy transport and in order for them to exist, the presence of a Dirac cone in the middle domain is crucial, which allows the strong coupling and hybridization of the two topological interface states associated with the middle domain due to its gapless bulk spectrum. One could expect that the coupling between the two interface states would become weaker if the width of the middle domain increases and consequently, the operational bandwidth of the topological waveguide states would become smaller. However, the scaling laws of the operational bandwidth as a function of the width of the middle domain in these systems are not known.

\begin{figure}
\includegraphics[width=\columnwidth]{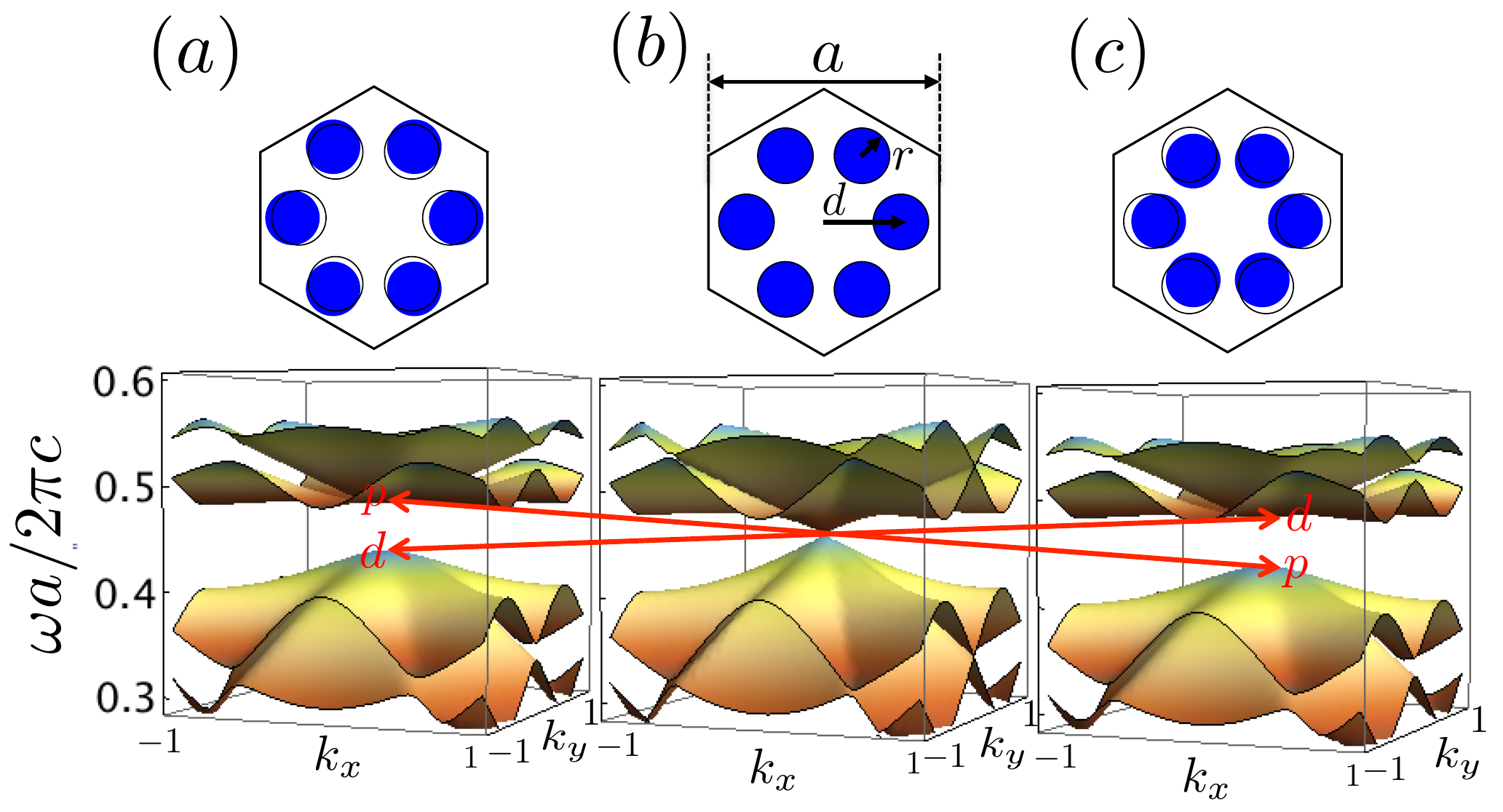}
\caption{Unit cells and band diagrams of the three photonic crytals for constructing the topological heterostructure. (a) The expanded unit cell and its topological nontrivial band diagram with a band inversion between the $p$ and $d$ orbitals. (b) The unit cell that gives a gapless band diagram with a double Dirac cone at $\Gamma$. (c) The shrunken unit cell and its trivial band diagram. The band inversion is indicated by the red arrowed lines. As labeled in (b), the lattice constant, radius of the  dielectric cylinder, and distance of the cylinder to the unit cell center are $a$, $r$ and $d$ respectively. Here, $r=0.12a$, $d=0.36a, a/3, 0.3a$ for (a-c) and the dielectric constant of the cylinder $\epsilon=12$.  \label{fig:fig1}}
\end{figure}

As photonic quantum spin Hall effect is an important class of phenomena in topological photonics, it is natural to ask whether large-area topological waveguide states could also be created utilizing this effect. In this work, taking the Wu-Hu model \cite{QSH_15PRL_Wu} widely used for creating pseudospin-momentum-locked helical electromagnetic edge states, we show that inserting a domain of photonic crystal with a double Dirac cone dispersion into two domains of photonic crystals with expanded and shrunken unit cells,  a pair of large-area helical waveguide modes emerges within a finite bandgap after gapping out the double Dirac cone in the middle domain. Importantly, we show that the operational bandwidth of these large-area modes decays as a power law with respect to the width of the middle domain and pseudospin-momentum locking unidirectional propagations as well as robustness of these large-area waveguide modes against sharp bends are further explicitly demonstrated. Considering that the Hu-Wu model has already found many interesting applications, such as coupling with quantum emitters \cite{Barik18Science_quantum}, reconfigurable devices \cite{Shalaev18NJP_reconfig,Cao19SB_reconfig}, all-optical logic gates \cite{He19OE_logic}, third-harmonic generation \cite{Smirnova19PRL_THG}, topological lasing \cite{Shao20NatNanoT_laser,Yang20PRL_laser}, and bound topological edge state in the continuum \cite{Zhang21PRApp_BIC}, the demonstration of large-area waveguide modes in this setup could offer new opportunities benefitting from the additional width degree of freedom, e.g., high-capacity topological transport and easy-interfacing with conventional waveguides and devices.

\begin{figure}
\includegraphics[width=\columnwidth]{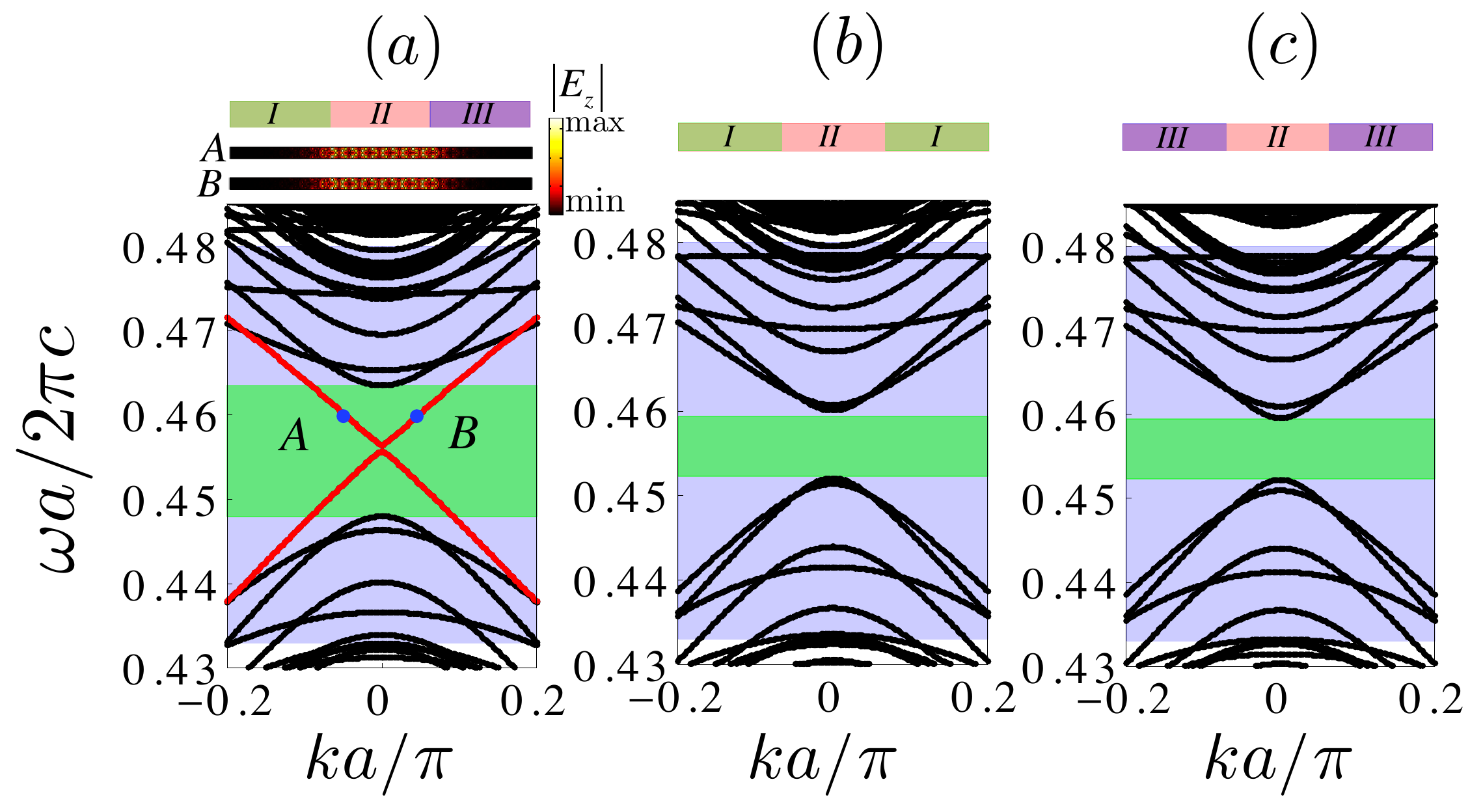}
\caption{(a) Projected band diagram of the three-layer topological heterostructure I/II/III, calculated by imposing periodic boundary conditions along the vertical direction whereas scattering boundary conditions along the horizontal direction of the supercell. The three domains I, II and III are constructed from the unit cells of (a), (b) and (c) in Fig.(\ref{fig:fig1}) respectively, and the numbers of the unit cells along the horizontal direction in the three domains are $N_I=N_{II}=N_{III}=10$. The red lines are the large-area helical waveguide modes within a bandgap indicated by the green region, whereas the blue region shows the common bulk bandgap of domains I and III. The mode profiles of two modes labeled by A and B are also shown. (b) and (c) Projected band diagrams similar to (a) but for two ordinary three-layer heterostructures with configuration of I/II/I and III/II/III, which can not support the large-area helical waveguide modes as in (a).  \label{fig:fig2}}
\end{figure}

\textit{\it Three photonic crystals for constructing the topological heterostructure.---} 
We begin by briefly discussing the Wu-Hu model \cite{QSH_15PRL_Wu} for emulating pseudospin-momentum-locked helical photonic edge states based on the $C_{6v}$ crystalline symmetry. The model considers photonic crystals with six dielectric cylinders in a hexagon unit cell (see Fig.\ref{fig:fig1}). When the distance of the cylinders to the unit cell center $d=a/3$, where $a$ is the lattice constant, the cylinders reduce to a honeycomb array and due to the band-folding mechanism, the original Dirac points at $K/K'$ associated with the transverse magnetic modes of the honeycomb lattice are folded into a double Dirac point at the $\Gamma$ point (see Fig.\ref{fig:fig1}(b)). Starting from this photonic crystal with a double Dirac point, a topological nontrivial (or trivial) photonic crystal with a finite bandgap could be created by  expanding (or shrinking) the six cylinders away from  (or towards) the unit cell center, see Fig.\ref{fig:fig1}(a) (or Fig.\ref{fig:fig1}(c)). Due to the $C_{6v}$ symmetry of the cylinders, the $\{p_x,p_y\}$ and $\{d_{xy},d_{x^2-y^2}\}$ orbitals associated with the two two-dimensional irreducible representations of the  $C_{6v}$ point group at the $\Gamma$ point form two double-degenerate pairs and a band inversion could be induced by this shrinking-expanding operation (see the red arrows in the band diagrams of Fig.\ref{fig:fig1}). Especially, the $p$ orbitals will be higher in frequency than the $d$ orbitals in the expanded unit cell, resulting in a topological nontrivial bandgap whereas the shrunken unit cell gives a topological trivial bandgap. While the original proposal of the Wu-Hu model only uses two photonic crystals made of expanded and shrunken unit cells for creating helical edge states along their interface, in the following, we will show that inserting a photonic crystal with the double Dirac cone into this interface of two photonic crystals with expanded and shrunken unit cells can lead to the emergence of large-area pseudospin-momentum-locked helical waveguide states.

\begin{figure}
\includegraphics[width=\columnwidth]{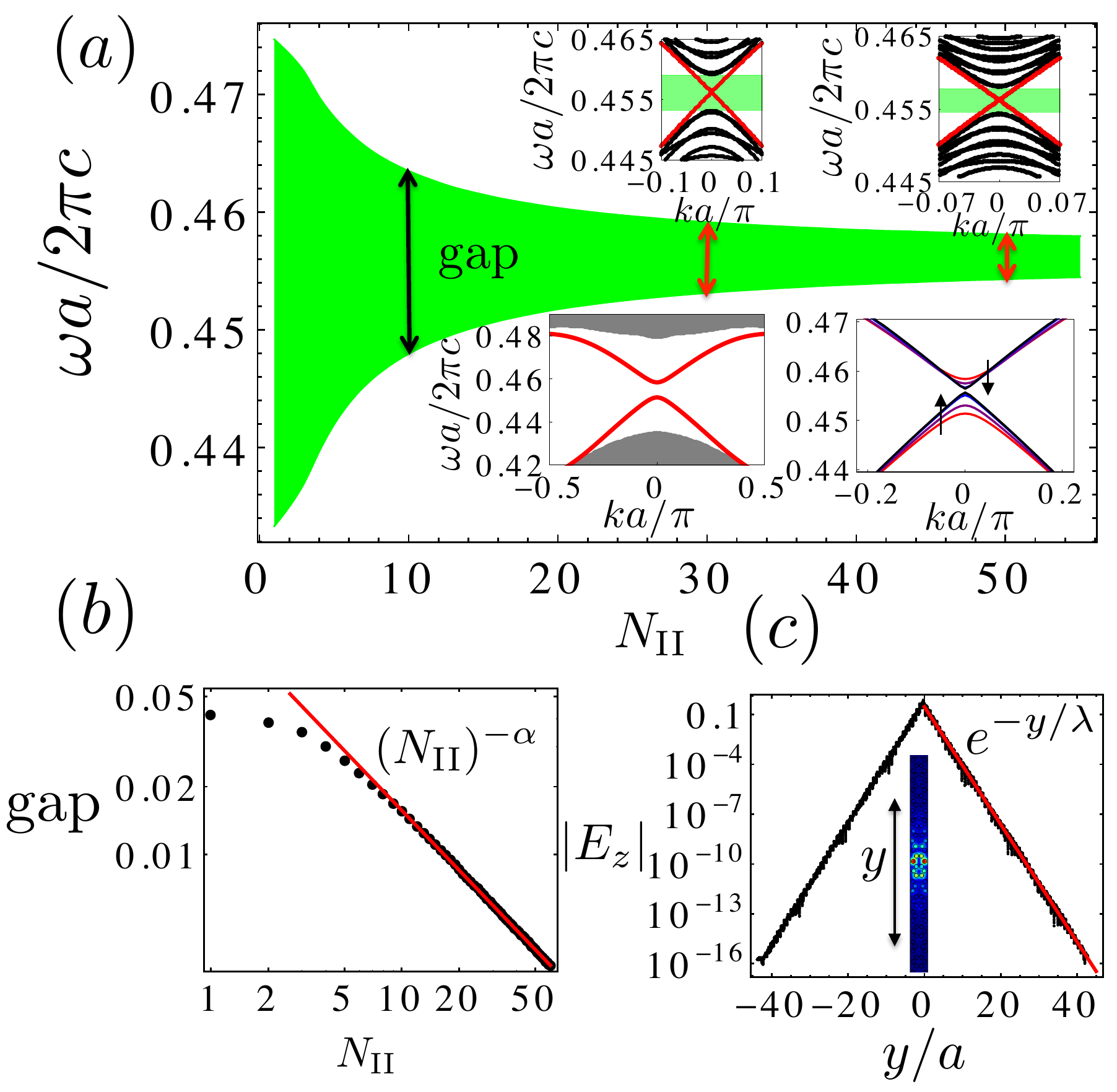}
\caption{(a) The gap within which the large-area helical waveguide modes reside as a function of the number $N_{II}$ of the unit cells in the middle domain of the three-layer heterostructure while fixing $N_I=N_{III}=10$. The top two insets show the projected band diagrams at $N_{II}=30$ and $N_{II}=50$, respectively, whereas the bottom two insets show the quantum-spin-Hall edge states of the Wu-Hu model (left) and their transition to the large-area waveguide states (right) when increasing $N_{II}$ from 0 to 1 to 5 and to 10 along the direction indicated by the arrows. (b) The size of the bandgap indicated in (a) as a function of $N_{II}$ on a log-log scale, indicating a power law scaling, i.e., $\text{gap} \propto N_{II}^{-\alpha}$ with $\alpha \sim 0.88$. (c) Decay behavior of the conventional quantum-spin-Hall edge states without the middle domain (i.e., $ N_{II} =0 $), which shows an exponential law, i.e, $|E_z|$ (integrated over width $a$ along $x$ and normalized) $\sim e^{-y/\lambda}$ with $\lambda \simeq1.23a$ at $ka/\pi=0.15$. The inset shows the field distribution of the mode around the interface. 
 \label{fig:fig3}}
\end{figure}

\textit{Emergence of large-area helical waveguide modes.---} To demonstrate large-area waveguide modes with quantum spin Hall features could be created, we consider a three-layer heterostructure whose supercell (see Fig.\ref{fig:fig2}(a)) consists of three regions I, II and III, which are filled with the unit cells in Fig.\ref{fig:fig1}(a), (b) and (c) respectively. The projected band diagram of this topological heterostructure is shown in Fig.\ref{fig:fig2}(a), from which one can see the emergence of a pair of waveguide modes (the two red lines) exhibiting the typical helical feature of conventional quantum spin Hall edge states, i.e., at each frequency there are two modes with opposite group velocities and thus opposite propagation directions.

Furthermore, the frequency window within which only helical waveguide modes exist (see the shaded green area of Fig.\ref{fig:fig2}(a)) is not the same as the original bandgaps (see the shaded light blue area of Fig.\ref{fig:fig2}(a)) of the two photonic crystals in regions I and III. Due to the strong coupling of the interface states at the I/II and II/III interfaces, the helical waveguide states in the middle domain II share both the features of topological interface states at the I/II and II/III interfaces and the bulk states in the middle domain II. Apart from the helical waveguide states, there are other nontopological waveguide modes appearing within the bandgaps of the I and III domains (see the black modes within the shaded light blue areas in Fig.\ref{fig:fig2}(a)) due to the partially gapping out of the bulk states in domain II because of its finite width. To show that the waveguide modes associated with the two red branches in Fig.\ref{fig:fig2}(a) indeed have large-area, we select two representatives (modes A and B) and present their field distributions at the top of Fig.\ref{fig:fig2}(a), from which one can see that the field amplitudes are almost uniformly distributed in the middle domain and decay exponentially into the photonic crystals in regions I and III due to the bandgap effect of domain I and III.

To exclude the possibility that these large-area waveguide states are just the bulk states of the middle domain, we further study the projected band diagrams of two ordinary heterostructures with configuration of I/II/I and III/II/III, where the two domains at each side of the middle domain have the same topological property, i.e., the difference of their topological invariants across the middle domain is trivial. From the results shown in Figs.\ref{fig:fig2} (b) and (c), one can clearly see the absence of the gapless helical waveguide states, where only gapped conventional waveguide states exist. This unambiguously demonstrates that two domains with different topological properties are needed to sandwich the middle Dirac-cone domain in order for the large-area quantum-spin-Hall waveguide states to emerge, which agrees with the general expectation of bulk-edge correspondence principle in topological physics. Despite their large-area, these quantum-spin-Hall waveguide states have inherent chirality and exhibit pseudospin-momentum locking helical propagation (will be demonstrated later), which are different from both the one-way large-area topological waveguide states studied in \cite{Wang21PRL_large} and the large-area valley-locked waveguide states studied in \cite{Chen21ACSpho_largevalley}, whose propagation directions are locked to the different valleys. Due to their large-area controllable by the domain width, these helical waveguide states could be used for high-capacity energy transport, which are different from the conventional helical edge states highly localized around the domain-wall interface \cite{QSH_15PRL_Wu}.

\textit{Power-law scaling of the operational bandwidth with respect to the middle domain width.---} A special feature of this kind of large-area waveguide states is that their operational bandwidth would decrease when the width of the middle domain becomes larger since the coupling of the interface modes associated with the interfaces I/II and II/III would become weaker. However, the exact scaling laws of the operational bandwidth with respect to the width of the middle domain were not explored in previous works \cite{Wang21PRL_large,Chen21ACSpho_largevalley}. To understand this scaling law for the large-area helical waveguide modes in our designed heterostructure, we increase the width of the middle domain II while fixing the widths of domain I and III to have 10 unit cells (i.e., varying $N_{II}$ while fixing $N_I=N_{III}=10$) and present the operational bandwidth as a function of $N_{II}$ in Fig.\ref{fig:fig3}(a).  First, to show that the large-area waveguide states revealed in Fig.\ref{fig:fig2} (a) at finite $N_{II}$ are descended from the quantum-spin-Hall edge states of the Wu-Hu model (see the bottom left inset of Fig.\ref{fig:fig3} (a), where the gap between the two branches of quantum-spin-Hall edge states are due to the reduction of $C_6$ crystalline symmetry at the domain-wall interface as originally pointed out by Wu and Hu \cite{QSH_15PRL_Wu}) and thus have a topological origin, we study the fate of the Wu-Hu quantum-spin-Hall edge states when increasing $N_{II}$ from 0. The results in the bottom right inset of Fig.\ref{fig:fig3} (a) show that when $N_{II}$ increases from 0 to 1 to 5 and to 10 as indicated by the direction of the arrows, the quantum-spin-Hall edge states of the Wu-Hu model smoothly evolve into the large-area quantum-spin-Hall waveguide states and notably, the gap between the two branches of large-area waveguide states becomes smaller due to the fact that the middle domain II has structure parameters that interpolate between those of domain I and III, thus mitigating the crystalline symmetry mismatch.

The gap results in Fig.\ref{fig:fig3}(a) show that in general, the size of the topological frequency window decreases when $N_{II}$ increases. To show the scaling more explicitly, we plot the size of the topological frequency window versus $N_{II}$ on a log-log scale in Fig.\ref{fig:fig3}(b), from which a clear linear fit at large $N_{II}$ can be found, indicating a power law scaling (i.e., $\text{gap} \propto N_{II}^{-\alpha}$) with the exponent to be $\alpha \sim 0.88$. The power-law scaling implies that the width of the waveguide (i.e., the middle domain) can be as large as one likes, though at the expense of a smaller operational bandwidth. To see this, we present the band diagrams at much larger $N_{II}$ of $N_{II}=30$ and 50 as two insets in Fig.\ref{fig:fig3}(a), from which one can see that, indeed, apart from the diminishing bandwidth (i.e., the green area), the helical waveguide modes alway exist. In general, with the increasing of the middle domain width, the topological frequency window would become narrow as more and more nontopologial waveguide modes enter the bulk gap and one can expect that in the limit of infinite width of the middle domain, the topological frequency window would eventually disappear.

To highlight the unique feature of the power law scaling, it may be enlightening to contrast it with the exponential decay law of conventional topological interface modes away from the interface without the middle domain. Fig.\ref{fig:fig3}(c) shows the decay of $|E_z|$ as a function of distance away from the interface, which clearly gives an exponential law, i.e., $|E_z|\sim e^{-y/\lambda}$ with the penetration depth $\lambda\simeq1.23a$ at $ka/\pi=0.15$. While the penetration depth of the interface modes in general depends on $k$ due to the bandgap effect, i.e., these locating deep within the bandgap have a shorter penetration depth compared to the modes close to the bandgap edge, its typical scale is on the order of lattice constant $a$. The existence of these two contrasting scaling laws is due to their different underlying physics, i.e., while the exponential decay of the conventional topological interface modes away from the interface originates from the bandgap effect, the power-law scaling of the operational bandwidth comes from the gapping out of the bulk states around the Dirac cone in the middle domain due to its finite width and the mode interaction.

\begin{figure}
\includegraphics[width=\columnwidth]{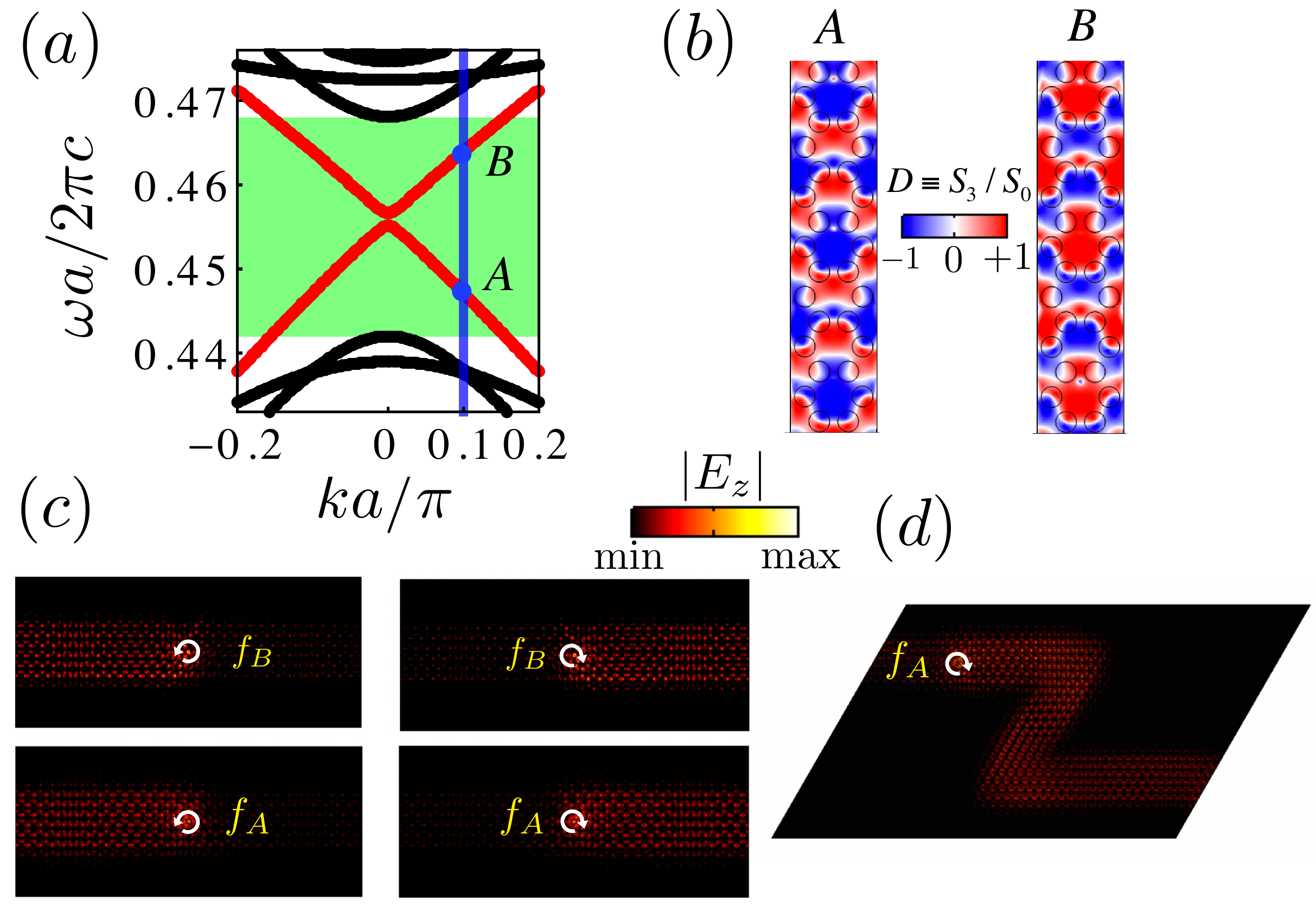}
\caption{Topological properties of the large-area helical waveguide modes. (a) Projected band diagram at $N_{II}=5$. (b)  Chirality (or directionality) map in the middle domain for the two modes A and B labeled in (a). (c) Pseudospin-momentum locking unidirectional propagations of the large-area helical waveguide modes. (d) Robustness of the topological large-area helical modes against sharp bends. The chirality of the excitation source is indicated by the clockwise or anticlockwise arrow. \label{fig:fig4}}
\end{figure}

\textit{Pseudospin-momentum locking unidirectional propagations against sharp bends.---}  The large-area helical waveguide modes in our designed heterostructure exhibit topological features, such as pseudospin-momentum locking propagations and possess inherent chirality, which could be exploited to realize unidirectional excitations and propagations. To show this, the mode dispersion of the helical waveguide states with $N_{II}=5$ is shown in Fig.\ref{fig:fig4}(a), where two modes marked by A and B at $k=0.1\pi/a$ (indicated by the blue line) are chosen for demonstration. The local chirality (or directionality) of the waveguide modes could be characterized by the Stokes parameters defined for the magnetic field $\mathbf{H}=(H_x,H_y)$ as $D=S_3/S_0$, where $S_0=|H_x|^2+|H_y|^2$ and $S_3=-2\textrm{Im}(H_xH_y^*)$ \cite{Lan21PRA_valleySHG}. The chirality maps in the middle domain II  of the heterostructure for the two modes A and B are presented in Fig.\ref{fig:fig4}(b), from which one can see that they exhibit nontrivial distributions with clear positive and negative chiralities. Furthermore, the chirality map of mode A is opposite to mode B, which is consistent with their opposite group velocities.

Employing the chirality map, pseudospin-momentum locking unidirectional propagation can be realized by using either left- or right-circularly polarized excitation source (in the simulations, six dipoles with phase winding along the clockwise or anti-clockwise direction are used as the circularly polarized excitation source). The results of the full-wave simulations for unidirectional excitations and propagations are presented in Fig.\ref{fig:fig4}(c), where the location and chirality of the source is indicated by the white arrow. The top (bottom) two panels of Fig.\ref{fig:fig4}(c) show that the large-area waveguide modes at frequency $f_B$ ($f_A$) indeed show pseudospin-momentum locking unirectional propagations.  Note that both the opposite chiralities and group velocities of modes A and B indicate that the same right- or left-circularly polarized source can excite the large-area waveguide modes at both frequencies $f_A$ and $f_B$ along the same direction. The robustness of the unidirectional propagations is topologically protected by the $C_{6v}$ crystalline symmetry. To demonstrate this, two sharp bends are introduced to the middle domain of the heterostructure and as can be seen from the full-wave simulation result shown in Fig.\ref{fig:fig4}(d), the unidirectional propagation of the large-area mode can pass the two sharp bends without being backscattered. We would like to note that while the conventional quantum spin Hall edge states confined at the interface between two photonic crystals show some robustness against disorder or defect, this immunity is not universal, i.e., it crucially depends on the type and location of the disorder or defect. Thus the robustness of the unidirectional propagation of large-area helical waveguide states against such kind of disorder or defect will not be further explored here.

\textit{Conclusion and outlook.---} In conclusion, we have investigated a new kind of large-area helical waveguide states in a three-layer topological photonic crystal heterostructure exploiting the photonic analogs of quantum spin Hall effect. These large-area waveguide states distribute uniformly in the middle domain of the heterostructure and exhibit pseudospin-dependent unidirectional propagation. Importantly, the size of the bandgap within which these large-area states reside shows a power-law dependence on the width of the middle domain, implying these large-area states in principle could exist in the middle domain with arbitrary width. The topological features of these large-area states, such as pseudospin-momentum locking unidirectional propagations and robustness of these states against sharp bends, have been explicitly demonstrated. 

The large-area helical waveguide modes studied in this work could readily be confirmed by inserting a domain of Dirac-cone PhC into the experimental setup of \cite{Yang18PRL_spinExp}, where helical  interface modes between two domains of topologically trivial and nontrivial PhCs have been observed in experiments. Our work could also open up other interesting directions for investigation. For example, up to now, the middle domain used in the topological heterostructure for demonstrating large-area waveguide states all features point degeneracies (Dirac points) and whether photonic structures featuring line degeneracies could also be utilized as the middle domain is also worthy of investigation. Furthermore, creating multiband larges-area waveguide states by combining both the quantum valley Hall and quantum spin Hall effects could further enrich possible practical applications \cite{Chen20PRR-spinvalley,Wei22PhoRes-spinvalley}. The fate of these large-area helical waveguide states under the effect of gain and loss is also an interesting problem \cite{Song20PRL_gainloss,Hou21PRB_gainloss,Chen21PRA_gainloss}. Finally, understanding how to tune the exponent of the power law and thus to control the bandwidth of the large-area waveguide modes in different scenarios is also important for both fundamental and practical interests. 

\textit{Acknowledgments.---} J. W. Y. acknowledges the support by the National Natural Science Foundation of China (62101124), Natural Science Foundation of Jiangsu Province (BK20210209), and the 111 Project (111-2-05); W. E. I. S. acknowledges the support by the National Natural Science Foundation of China (Nos. U20A20164 and 61975177).

\end{document}